# Neutronics Analysis for MSR Cell with Different Fuel Salt Channel Geometry

Authors: Shi-he Yu[1]　Ya-fen Liu[1,*]　Pu Yang[1]　Rui-min Ji[1]　Gui-feng Zhu[1]　Bo Zhou[1]　Xu-zhong Kang[1,2]　Rui Yan[1]　Yang Zou[1]　Ye Dai[1]

**Affiliations:**

[1]Shanghai Institute of Applied Physics, Chinese Academy of Sciences, Shanghai 201800, China

[2]University of Chinese Academy of Sciences, Beijing 100049, China

[*]Corresponding author. *E-mail address:* liuyafen@sinap.ac.cn

**Abstract:** The neutronic properties of Molten Salt Reactor (MSR) are different from that of traditional solid-fuel reactors due to its nuclear fuel particularity. Based upon MCNP code, the influence of the size and shape of fuel salt channel on neutron physics of MSR cell was studied systematically in this work. The results show that the infinite multiplication factors ($k_\infty$) increases first and then decreases with the change of graphite cell size under the condition of given fuel volume fraction (FVF). In the case of the same FVF and average chord length, when the average chord length is relatively small, the $k_\infty$ values with different fuel salt channel shapes are in good agreement; when the average chord length is relatively large, the $k_\infty$ values with different fuel salt channel shapes are greatly different. In addition, some examples of practical application of this work are illustrated in the end, including cell selection for the core and thermal expansion displacement analysis of the cell.

**Introduction:** As one of the six Gen-IV reactor types, molten salt reactor (MSR) was first studied in 1950s as a means of providing a compact high temperature power plant for nuclear powered aircraft [1]. Aircraft Reactor Experiment (ARE) was then constructed at Oak Ridge National Library (ORNL), USA, which demonstrated the nuclear feasibility of operating a molten salt fueled reactor at high temperature [2]. And after realizing the high promise of the molten salt reactor type for achieving low electric power generating costs in central power plant, Molten Salt Reactor Experiment (MSRE) was constructed and ran successfully [3]. In addition, Critical Experiment Device (hereafter the device) was established in Shanghai Institute of Applied Physics (originally called Shanghai Institute of Nuclear Research), China, in 1970s. Its main purpose was to explore the physics characteristics of MSR and thorium-uranium cycle. Subsequently, although the molten salt reactor research has not been interrupted, but did not reach the actual level of building reactor [4-6].

After rediscovering that molten salt reactor has unique advantages in inherent safety, economy, sustainable development of nuclear fuel and nonproliferation, the research of molten salt reactor has set off a new upsurge since the 21$^{st}$ century [7-9]. Aiming at the national major strategic needs in the field of nuclear energy, the Chinese Academy of Sciences (CAS) launched the strategic and pioneering science and technology project of "Future Advanced Fission Energy-Thorium based Molten Salt Reactor nuclear energy system" in 2011 and Thorium Molten Salt Reactor (TMSR) center was established subsequently. It is dedicated to the research and development of the Gen-IV advanced fission reactor nuclear energy system to achieve strategic goals such as diversifying nuclear fuel, preventing nuclear proliferation and minimizing nuclear waste. The preliminary target of the project is to build a 2 MW Thorium based Molten Salt Reactor (TMSR-LF1) [10, 11].

Molten salt reactor is a kind of thermal neutron reactor that takes use of molten salt mixture fuel cycle and moderated by graphite to produce fission energy. Molten salt mixture fuel can be fluoride of uranium, thorium, zirconium, etc., which is used as both nuclear fuel and heat carrier under high temperature liquid state without the need for solid fuel element [12]. In general, the physical characteristics of the core are studied by changing the volume ratio between the moderator and the nuclear fuel in the thermal reactors, such as the Pressurized Water Reactors (PWRs) [13, 14]. For molten salt reactor, its lattice cell is also the component of traditional reactor. The cells are usually columnar graphite blocks, and the side length increases with the increase of the core size. For example, the side length of regular quadrilateral graphite lattice cell of MSRE (Molten Salt Reactor Experiment) is 5 cm, while the side length of hexagonal graphite lattice cell of MSBR (Molten Salt Breeder Reactor) is 15 cm [15-18]. Even if two graphite cells with the same geometric shape and have the same volume ratio of graphite to fuel salt, the different size of the cells will change the neutron physical properties. The geometry of liquid nuclear fuel in molten salt reactors is so flexible that molten salt reactors can have fuel cells of various shapes. In addition, because the coefficient of thermal expansion between graphite and alloy is quite different, the change of core geometry due to high temperature should not only consider the thermal expansion of each component, but also consider the displacement of the graphite cell. New fuel salt channels can be formed between the cells after the graphite cells being displaced, and their geometric shapes are different from the original fuel salt channel [19]. Therefore, it is necessary to study and analyze the neutron physical properties of different graphite cell size and different fuel salt channel shape.

Based upon MCNP (Monte Carlo N Particle transport code), several kinds of graphite cell models of fuel salt channel with different sizes and geometric shapes were displayed in this work

[20, 21]. The influence of the size and geometric shapes of the fuel salt channel on the $k_\infty$ and the neutron energy spectrum of the cell was studied subsequently. In addition, the variation law of the $k_\infty$ with the average chord length of the fuel salt channel under different fuel salt volume ratio and heavy metal molar ratio was calculated and analyzed. Finally, some practical application examples of this study are illustrated, which can be used as a reference for the design of molten sale reactor.

**1 Models and Parameters:** Molten salt reactors take many forms, but usually liquid fuel salts flow through cores made of graphite with critical geometry. The heat is produced by the fission of the fuel salt and is carried away by the flowing fuel salt. Fuel is concentrated into blocks in an inhomogeneous reactor, such as cylinders, circles, spheres, sheets (planks), etc., which are placed in the moderator in a certain geometric form to form the core of the so-called lattice structure. The elementary units that make up a lattice structure are generally called cells [22]. The fuel salt channel in molten salt reactor corresponds to the fuel block in the traditional reactor. For the purpose of research, the variation trend of the infinite medium multiplication factor ($k_\infty$) of the cell was only considered in this work. Then the neutron physics law of the corresponding type of cell was obtained by the $k_\infty$ value analyzing. Therefore, there was no need to model the whole core. The $k_\infty$ is related to the size and shape of the fuel salt channel, and the influence of the size and shape of the fuel salt channel on the $k_\infty$ can be constrained by the average chord length of the fuel salt channel $L$. As shown in Formula (1), the average chord length of the fuel salt channel is defined as 4 times the volume of the fuel salt channel $V$ divided by the surface area of the fuel salt channel $S$ [22].

$$L = \frac{4V}{S} \quad (1)$$

As shown in Fig. 1, several basic fuel salt channel models in different shapes were established by MCNP6 for comparative analysis in this work, including cylinder, annular, sphere and flake [20, 21]. The central region of each cell in Fig. 1 represents fuel salt and the surrounding region represents graphite. Typically, molten salt reactors choose a cylindrical shape for the fuel channel, such as MSBR for breeding purpose. The fuel salt channel of MSRE, which is well known, is of runway type and can be regarded as a combination of cylinder and flake. After the displacement of graphite cell, the new fuel salt channel formed between cells is generally flake. The reason why the circular and spherical fuel salt channel is chosen is for the comparison of results. It should be noted that the fuel salt can also flow up and down by gouging out a channel above and below the sphere. The cell is divided into fuel salt zone and graphite moderator zone. For example, in the cylindrical fuel salt channel model, the fuel salt zone is the cylindrical zone, and the graphite moderator zone is other regions outside the cylinder. Key parameters such as fuel salt composition and density used by the cell are shown in Table 1. The fuel salt composition is consistent with MSRE, so that the trend of density change with temperature can be obtained from MSRE technical report data [3]. The density of the fuel salt is 2.34 g/cm$^3$ at 550°C. T220-graphite was used in this work with a density of 1.86 g/cm$^3$ at room temperature and a linear thermal expansion coefficient of 3.49×10$^{-6}$ /°C at room temperature to 550°C [3]. The average temperature of the cell is 550°C, which was used in the calculation and analysis of this work. Nuclear data library ENDF/B-VII.0 was selected in the calculation of this work and its applicability and advantages to molten salt reactors were verified by Liu Yafen et al in 2018 [5, 23]. For the purpose of comparing results, all the calculations were performed under the same condition that the source neutrons selected for each cycle were 100 000, with 1050 active cycles and 50 abandoned cycles, leading to a standard deviation of about 0.00005.

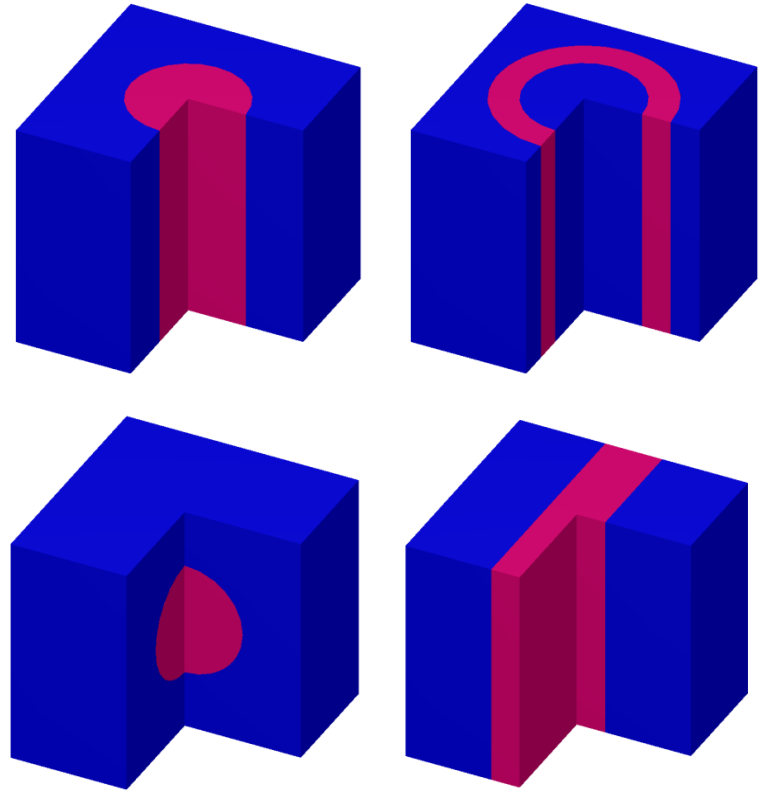

**Fig.1**. The Schematic diagrams of the computational models.

**Table 1.** Parameters of graphite cell model.

| Parameter | Value |
| --- | --- |
| Fuel salt composition, mole | 65.39%LiF-25.34%BeF$_2$-4.72%ZrF$_4$-1.55%UF$_4$ |
| Fuel salt density at 550°C, g/cm$^3$ | 2.34 |
| $^{235}$U enrichment, % | 17 |
| Graphite density at 550°C, g/cm$^3$ | 1.86 |
| Linear thermal expansion coefficient of graphite at room temperature to 550°C, /°C | 3.49×10$^{-6}$ |
| Linear thermal expansion coefficient of Hastelloy-N at room temperature to 550°C, /°C | 13.40×10$^{-6}$ |
| Fuel salt temperature, °C | 550 |

Even if the fuel salt fraction and the Fuel Volume Fraction (FVF) is the same, the $k_\infty$ of cell is related to the arrangement of the cell, which is not taken into account here. When changing the average chord length, only the relatively convenient method is adopted in this work, and the difference of the $k_\infty$ caused by other method is not considered. Method is shown in detail as follows: the average chord length of the cylindrical and spherical fuel salt channel was changed by changing the size of the channel; the average chord length of the circular fuel salt channel was changed by changing the outer diameter of the circular fuel salt channel. In addition, and the average chord length of the flaky fuel salt channel was changed by changing the thickness of the flaky fuel salt channel. The FVF of the cell was kept constant when the average chord length was generally changed in the cases above. The influence of the size and geometric shapes of the fuel salt channel on the $k_\infty$ and the neutron energy spectrum of the cell was studied firstly. Then, the variation law of the $k_\infty$ with the average chord length of the fuel salt channel under different FVF values and heavy metal molar ratio was calculated and analyzed. At last, some practical application examples of this study are illustrated.

## 2 Results and Discussion

**2.1 Influence of Fuel Salt Channel Geometry:** As described in the previous section, the fuel salt channel geometric shape considered in this section includes: cylindrical, circular, spherical, and flaky. The influence of different geometric shapes on $k_\infty$ and neutron energy spectrum of the cell was studied and analyzed in this section. In addition, the physical causes of these influences are analyzed together with the calculation results. Table 2 shows the graphite cell side length of fuel salt channel with different geometric shapes under the same average chord length. The data in the table is also an important input parameter for the subsequent calculation in this work. It can be considered that under the same average chord length, the graphite cell side length of fuel salt channel with different shapes differs greatly.

**Table 2**. Side length of cell with different fuel salt channel geometric shape.

| $L$, cm | 1.00 | 2.00 | 4.00 | 8.00 | 16.00 | 32.00 | 40.00 |
|---|---|---|---|---|---|---|---|
| Side length of cell with cylindrical fuel channel, cm | 2.29 | 4.58 | 9.15 | 18.31 | 36.60 | 73.20 | 91.02 |
| Side length of cell with flaky fuel channel, cm | 3.83 | 7.67 | 15.33 | 30.67 | 61.33 | 122.67 | 153.33 |
| Side length of cell with spherical fuel channel, cm | 2.28 | 4.55 | 9.11 | 18.20 | 36.41 | 72.80 | 91.00 |
| Side length of cell with circular fuel channel, cm | 3.96 | 7.92 | 15.85 | 31.70 | 63.40 | 126.79 | 158.49 |

Figure 2 shows the variation of the $k_\infty$ with the average chord length of fuel salt channel under different fuel salt channel geometric shapes. It can be seen from the figure that when the average chord length is relatively small ($L<7.5$ cm), the $k_\infty$ of fuel salt channel with different geometric shapes accords well with each other. This is because when the same $L$ value is less than 7.5 cm, the energy spectrum of the fuel region under different geometric shapes is almost the same. As the average chord length increased to 7.5 cm, the difference of the $k_\infty$ of fuel salt channel with different geometric shapes begins to be significant. In addition, in the relatively large range (7.5 cm~40 cm) of average chord length, $k_\infty$ of cylindrical and spherical fuel salt channel is greater than $k_\infty$ of flaky or circular fuel salt channel. Comparatively speaking, the $k_\infty$ of flaky and circular fuel salt channel is close to each other. The $k_\infty$ of cylindrical fuel salt channel is similar to that of spherical fuel salt channel.

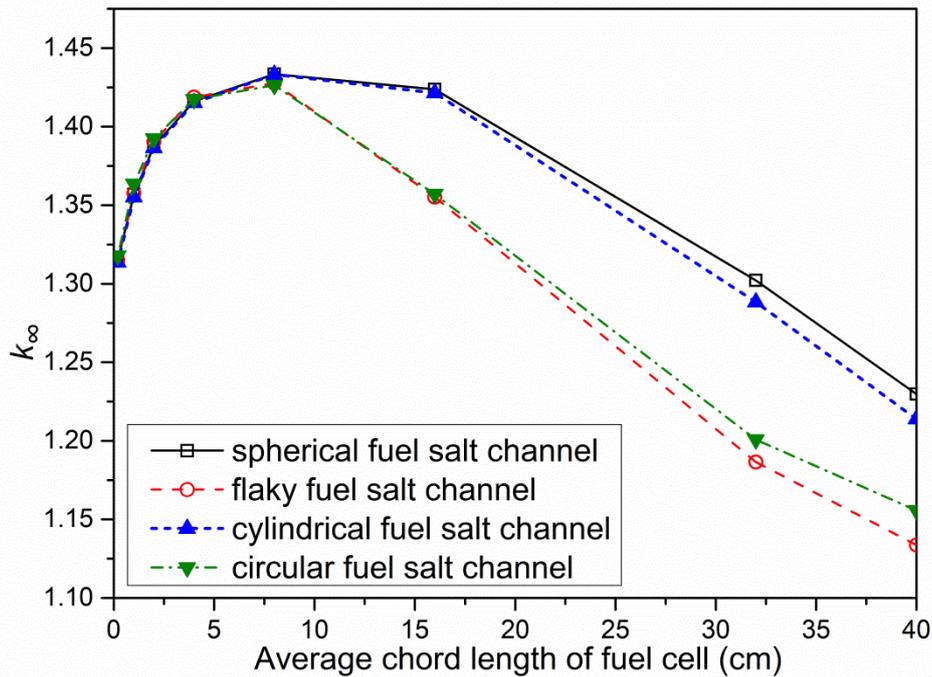

**Fig.2**. $k_\infty$ at different chord lengths for different fuel salt channel shape.

Figure 3 shows the normalized neutron flux distributions of the cell (left) and fuel salt zone (right) with different geometric shapes. The FVF of each cell was 15%, and the average chord length was uniformly set as 16 cm. It can be seen from the figure that the neutron spectrum of the cell with fuel salt channel shape of flake and circular is very consistent. In addition, the neutron spectrum of the cell with cylindrical fuel salt channel is similar to that of the cell with spherical fuel salt channel. As can be seen from the normalized neutron flux distribution in cell (the left of Fig. 3), the neutron spectrum of the cell with flaky and circular fuel salt channel is relatively thermal and the thermal neutron utilization factor is comparatively small. Furthermore, the normalized neutron flux distribution in fuel salt (the right of Fig. 3) indicates that the neutron spectrum of the flaky and circular fuel salt channel is relatively hard and the neutron share in the resonance absorption region also increases correspondingly. Therefore, in the interval with relatively large average chord length, the variation of $k_\infty$ appears as shown in Fig. 2.

Through the above calculation and analysis, it can be found that although the FVF and FC of the cell are the same, the geometric shape of fuel and salt channel is different, which has a great influence on the neutron spectrum, leading to a great discrepancy in the $k_\infty$. The average chord length has a certain constraint on the $k_\infty$. When the average chord length is relatively small (less than 10 cm), the $k_\infty$ of fuel salt channel with different geometric shapes is in good agreement. In the interval with relatively large average chord length (greater than 10 cm), the $k_\infty$ variation trends of fuel salt channel with different geometric shapes are the same, but the specific $k_\infty$ values differs greatly. The next section will give the detailed reasons for these $k_\infty$ differences.

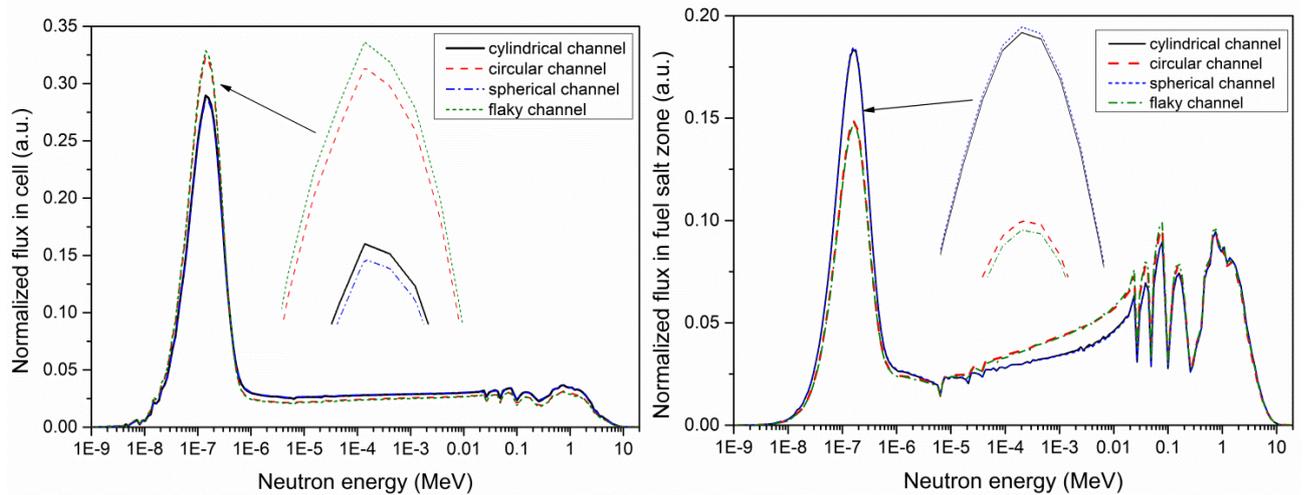

**Fig.3**. Neutron spectrum of cell (left) and fuel salt zone (right) for different fuel salt channel shape with $L$=16 cm.

**2.2 Influence of Fuel Salt Channel Size:** The graphite cell with circular fuel salt channel was selected for calculation and analysis in the work of this section. Under the condition that the material composition of cell is determined and FVF value is fixed at 15%, the size of the cell was changed to calculate the variation of $k_\infty$. The variation curve of this value is shown in Table 3 and the emphasis of this section is to explain the reason of the change from the four factors and the neutron energy spectrum of cell. The calculation method of the four factors was shared by Sun Jianyou et al in 2014 [24, 25]. And the results of these factors are also displayed in Table 3, where ε is the fast neutron multiplication factor; $p$ represents the probability of escape resonance absorption; η is the effective number of fission neutrons and f represents the thermal utilization factor.

**Table 3.** Variations of the four factors with the radius of the fuel salt channel.

| Fuel salt channel radius, cm | 0.5 | 1.0 | 2.0 | 4.0 | 8.0 | 12.0 | 16.0 | 20.0 |
|---|---|---|---|---|---|---|---|---|
| $\varepsilon$ | 1.08291 | 1.08107 | 1.07896 | 1.07611 | 1.07600 | 1.10451 | 1.13528 | 1.18708 |
| $p$ | 0.78633 | 0.80635 | 0.82658 | 0.84465 | 0.85663 | 0.82149 | 0.78896 | 0.73737 |
| $\eta$ | 1.64413 | 1.64401 | 1.64504 | 1.64313 | 1.64264 | 1.64247 | 1.64239 | 1.64214 |
| $f$ | 0.96802 | 0.96729 | 0.96537 | 0.95946 | 0.93877 | 0.90875 | 0.87553 | 0.84484 |
| $k_\infty$ calculated based on the four-factor formula | 1.35525 | 1.38625 | 1.41633 | 1.43296 | 1.42136 | 1.35685 | 1.28797 | 1.21437 |
| $k_\infty$ calculated by MCNP6 | 1.35513 | 1.38642 | 1.41524 | 1.43321 | 1.42142 | 1.39675 | 1.28830 | 1.21383 |

As can be seen from Table 3, with the increase of fuel salt channel radius, the value of $k_\infty$ presents a trend of first increasing and then decreasing. The variation of the four factors is summarized as follows:

- $\varepsilon$ first decreases and then increases;
- $p$ first increases and then decreases;
- $\eta$ is basically unchanged with the change of fuel salt channel radius;
- And $f$ keeps decreasing with the increasing of fuel salt channel radius.

The results indicate that the variation of $k_\infty$ is mainly caused by the escape resonance absorption probability $p$ and thermal neutron utilization coefficient $f$. Although the fast neutron multiplication factor $\varepsilon$ has a certain effect on $k_\infty$ value, the amplitude of its variation is much smaller than that of the escape resonance absorption probability.

Figure 4 shows the normalized neutron flux distributions in cell (left) and fuel salt zone (right) with different fuel salt channel radii, and R is the radius of fuel salt channel. With the increase of fuel salt channel radius, the neutron spectrum of the cell tends to be thermal. This is due to the fact that as the fuel salt channel radius increases and the distance between the channels increases, the graphite area expands, increasing the probability that the fast neutron in the graphite slows to below the resonance energy. In other words, the share of thermal neutron flux in graphite increases, which leads to the decrease of thermal neutron utilization of cell. Resonance neutrons are mainly produced in graphite moderators and then incident on fuel salts. Due to the large resonance absorption cross section of the nuclei in fuel, the resonance neutrons is first absorbed by the nuclei on the surface of fuel salt channel and acts as a shield to the nuclei inside the fuel salt channel. ΔR is set as the resonant absorption surface of fuel salt channel and R as the radius of fuel salt channel. An increase in R will not make ΔR change, and the area where neutrons are not absorbed by resonance increases. That is, the spatial self-screen effect increases, while resonance absorption decreases and the probability of escaping resonance absorption increases.

However, from the right side of Fig. 4, it can be seen from the neutron flux distribution of fuel salt zone that the spectrum becomes harder as the radius of fuel salt channel increases. In particular, the spectrum has an obvious hardening trend when the radius of fuel salt channel is larger than 8 cm. This is mainly due to the fact that the fuel salt itself contains moderated materials, such as lithium and beryllium, etc. When the radius of the fuel salt channel increases, the probability of neutron reaction in the fuel salt increases, so that the probability of neutron slowing down to the resonance region in the fuel salt increases. In other words, as the resonance absorption of the neutron increases, the probability of escaping from the resonance absorption decreases.

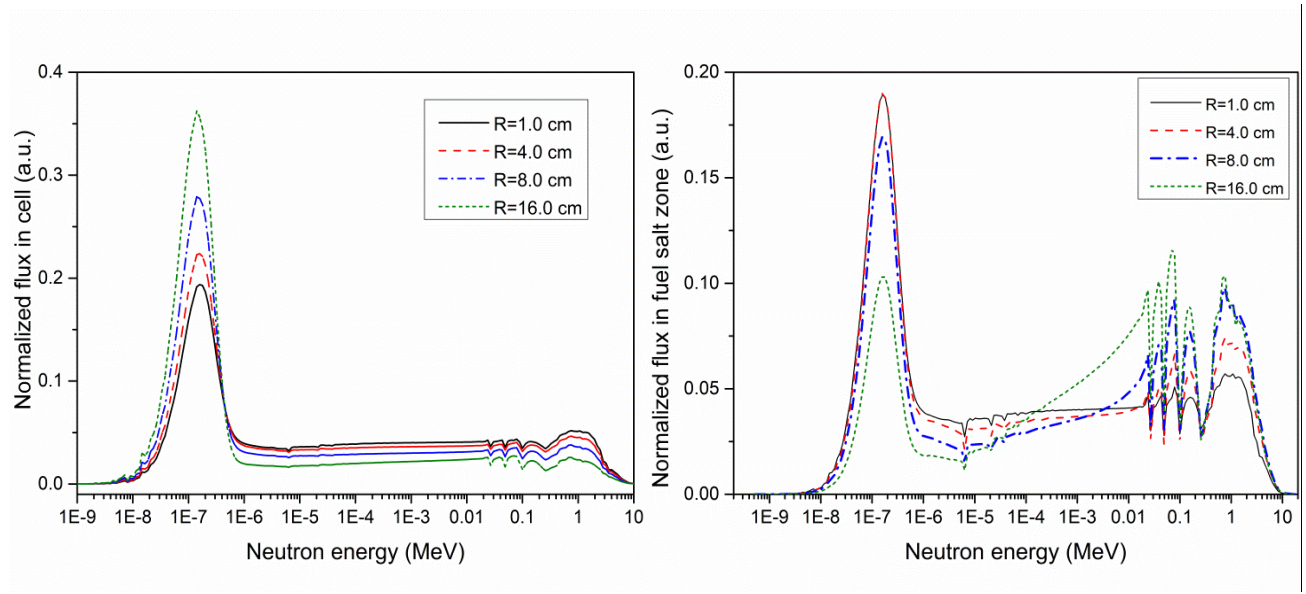

**Fig.4**. Neutron spectrum of cell (left) and fuel salt zone (right) under different fuel salt channel radii.

Based upon the above analysis, when the FVF of the cell with cylindrical fuel salt channel is fixed at 15%, the probability of escaping from the resonance absorption first increase and then decrease with the increasing of the fuel salt channel radius. The thermal neutron utilization coefficient always decreases with the increase of channel radius. Therefore, the overall effect is that as the fuel salt channel radius increases, the $k_\infty$ first increases and then decreases. It is worth noting that the fuel salt itself has the ability to slow down neutrons, therefore, the resonance neutrons are also produced in the fuel salt region, which has an effect on the resonance absorption of the cell.

**2.3 Influence of FVF and FC (Fuel Concentration):** The volume ratio of moderator to fuel and the composition of nuclear fuel change directly affect the physical properties of the core. The variation law of the $k_\infty$ value of cell was obtained under the given conditions of fuel salt volume ratio and fuel salt heavy metal molar ratio. This section focuses on the analysis of the variation law of the $k_\infty$ value with average chord length under different FVF and FC. The FVF value of cell was changed by adjusting the size of the graphite portion, while the size of the fuel salt channel remains unchanged. The value of FVF was set at 5%, 15% and 30%. The influence of FC on the $k_\infty$ value was also studied, and the FC values were set as 1.55%, 4% and 8%.

Figure 5 shows the $k_\infty$ variation of cell with cylindrical fuel salt channel under different FVF and FC with change of fuel salt channel radius. Although the $k_\infty$ was greatly affected by FVF, under different FVF conditions, this value increases first and then decreases with the change of fuel salt channel radius. In the case of FVF=5%, when the fuel channel radius is less than 1.5 cm, the $k_\infty$ value increases with the increase of the fuel channel radius. The $k_\infty$ value decreases with the increase of the channel radius when the fuel channel radius is larger than 1.5 cm. In other words, $k_\infty$ is maximized at a fuel channel radius of 1.5 cm. And $k_\infty$ is maximized at a fuel channel radius of 4.5 cm and 8 cm in the case of FVF=15% and 30% respectively. In the case of FC=1.55%, when the fuel channel radius is less than 4 cm, the $k_\infty$ value increases with the increase of the fuel channel radius. The $k_\infty$ value decreases with the increase of the channel radius when the fuel channel radius is larger than 4 cm. What`s more, $k_\infty$ are both maximized at a fuel channel radius of 7.5 cm in the case of FC=4% and 8% respectively. The $k_\infty$ variation trend of the fuel channel radius remains unchanged, only the radius of the fuel salt channel corresponding to the maximum $k_\infty$ value changes. This is mainly due to the different degrees of neutron slowing down under different FVF and FC results in different neutron energy spectra.

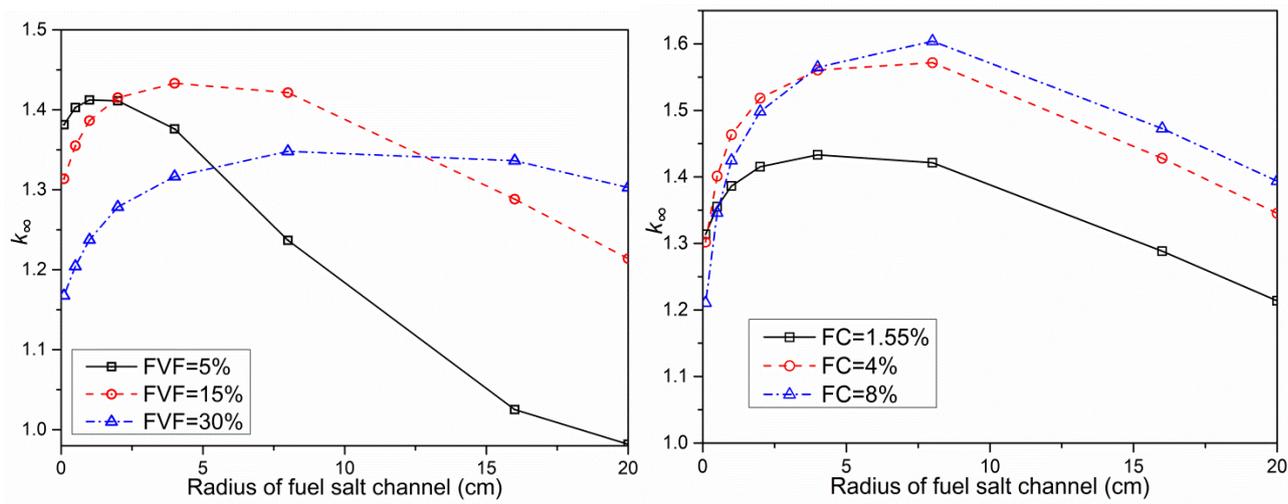

**Fig.5**. Effect of fuel salt channel radius on $k_\infty$ for different FVF (left) and FC values (right).

Figure 6 and 7 show the variation of $k_\infty$ of fuel salt channel with average chord length of fuel salt channel under different FVF and FC, respectively. PZ, Q, YZ and YH represent the flaky, spherical, cylindrical and circular fuel salt channel. It can be seen that the variation law of $k_\infty$ with average chord length is consistent under different FC values for the 4 fuel salt channels with different geometric shapes. In addition, the variation law of $k_\infty$ with average chord length is also consistent under different FVF values except for the case of circular fuel salt channel with FVF value 5%. Figure 8 shows the normalized neutron spectrum in cell of the circular fuel salt channel. In the region where FVF is low and the average chord length is large, due to the increase of the fast neutron multiplication coefficient, the spectrum becomes harder with the increase of the average chord length. The thermal neutron utilization coefficient increases correspondingly. Therefore, the $k_\infty$ has a tendency to become larger again.

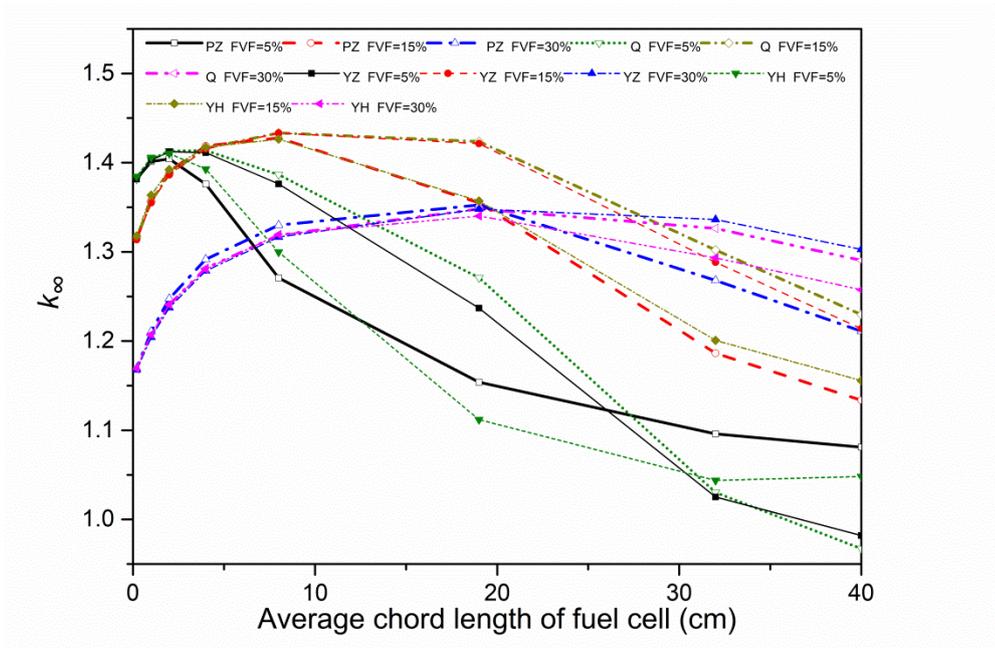

**Fig.6**. Effect of average chord length on $k_\infty$ for different FVF values.

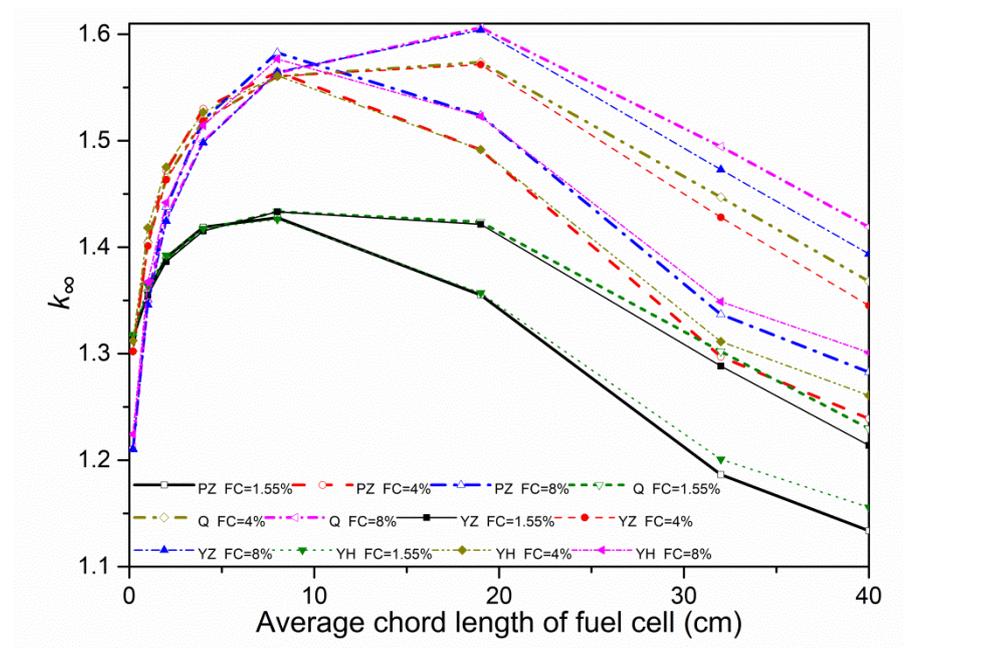

**Fig.7**. Effect of average chord length on $k_\infty$ for different FC values.

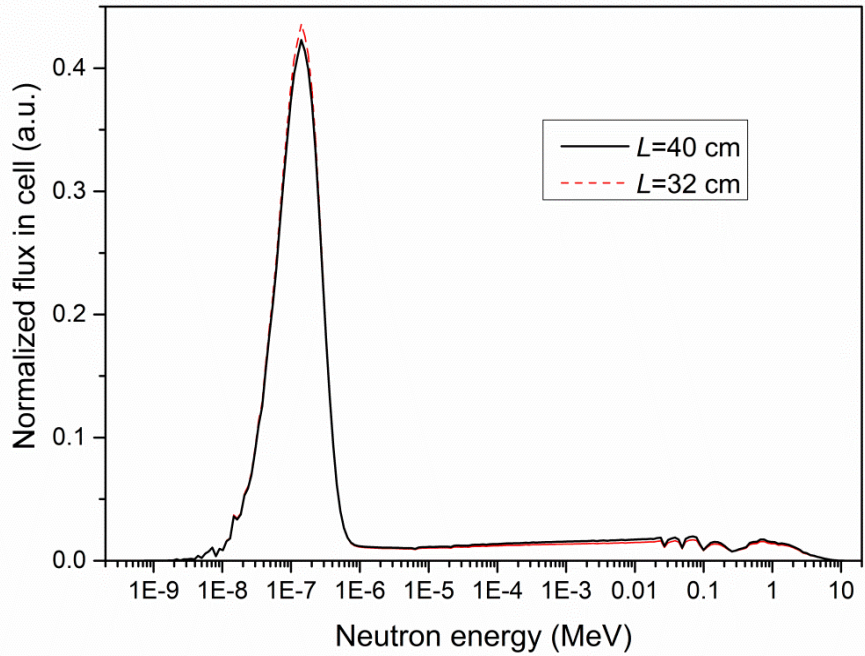

**Fig.8**. Normalized neutron spectrum of the circular cell.

Similarly, for the fuel salt channel cells with the same FVF and FC, $k_\infty$ of the fuel salt channels with different geometric shapes is in good agreement in the interval with average chord length less than 10 cm. The $k_\infty$ of the fuel salt channel with different geometric shapes is greatly different in the range of the average chord length greater than 10 cm. Moreover, the variation law of $k_\infty$ with average chord length of the cell with flaky and circular fuel salt channel is similar, while that of the cell with cylindrical and spherical fuel salt channel is similar. In summary, although FVF and FC are different, the variation law of the $k_\infty$ of cell with average chord length is universal.

## 3 Application Examples

It is well known that it is necessary to choose the size and shape of the grid element in the research and design of the actual molten salt reactor. Molten salt reactor cores can be large or small, and large graphite cells can be used for large cores to enhance the cell mechanical structure strength. At the same time, the characteristics of molten salt reactor itself, namely the high temperature state and the flow properties of fuel salt, determine the geometric diversity of fuel salt channel in the core. Therefore, the study of the influence of different flow passage geometry on neutron physics can provide reference for the selection of graphite cell.

**3.1 Cell Selection:** The fuel assembly in the molten salt core is the graphite cell in the core. According to the calculation data in Table 3, when FVF is fixed, that is, when the fuel salt load of the core is the same, there will be a maximum $k_\infty$ value with the change of the size of the graphite cell. Therefore, in the case that the FVF value is determined, the size of the graphite cell can be reasonably selected if only $k_\infty$ is considered in the optimal design of the cell.

It can be seen from Fig. 2 that the fuel salt channel with different geometric shapes has the same $k_\infty$ under the same FVF value when the average chord length is relatively small. As can be seen from Table 2, under the same average chord length, the graphite cell side length of fuel salt channel with different geometric shapes varies greatly. Therefore, under the condition of equal $k_\infty$, the length of the graphite cell with flake and circular fuel salt channel is much larger than that with cylinder and spherical fuel salt channel. And flaky or circular fuel salt channels can be used when larger graphite cells are needed.

**3.2 Effect of Expansion Displacement on Cell:** In molten salt reactor, the graphite component is located on the alloy support plate. There is a big difference in the expansion coefficient between the graphite and the alloy, which makes the spacing between the graphite gate elements more obvious at high temperature, thus graphite cell displacement is generated. Due to the fluidity of fuel salt, the gap between the graphite cells is filled with fuel salt after graphite cell displacement. Therefore, after the expansion and displacement of graphite cells, not only FVF changes, but a new fuel salt channel is formed between the cells, and the geometric shape is different from the original fuel salt channel. These changes will have a certain effect on the physical properties of the core.

It is assumed that the reference fuel salt channel is cylindrical and its FVF value was 15%. The second row in Table 4 displays the $k_\infty$ values of this reference cell with different chord lengths. As the temperature rises from 550 °C to 700 °C (accident temperature), the graphite cells will shift due to the difference in thermal expansion coefficient between graphite and Hastelloy-N. Areas that increase due to differences in coefficient of thermal expansion are filled with fuel salt. As a result, FVF increased from 15% to 15.4% and flaky new fuel salt channel appeared around the graphite cell. The line 5 in Table 4 represents the deviations between the $k_\infty$ values of the cell after displacement and the corresponding reference $k_\infty$ values. In order to distinguish the influence of FVF on the neutron physical properties of cell, the fuel salt channel radius was kept unchanged, and the side length of graphite gate element was artificially reduced to obtain the $k_\infty$ values with FVF increasing to 15.4%. The line 4 shows the differences between these $k_\infty$ values and the $k_\infty$ values of the corresponding cells after displacement. In summary, the difference in the third row in Table 4 reflects the impact of FVF increase on $k_\infty$. The difference in line 4 reflects the effect of different cell geometry (cylindrical and flaky) of new fuel salt channel on $k_\infty$.

As can be seen form the comparison results in Table 4, FVF changes and newly formed fuel salt channel caused by expansion displacement of graphite cell will both have an impact on $k_\infty$. In addition, the influence of the two on $k_\infty$ of the cell is also changed under different sizes of the cell. Therefore, the influence of material expansion displacement should be considered in the core design. Alternatively, the cell could be bound to prevent the cell from moving with the support plate.

**Table 4.** Effect of different factors on $k_\infty$.

| $L$, cm | 1.00 | 2.00 | 4.00 | 8.00 | 16.00 | 32.00 | 40.00 |
|---|---|---|---|---|---|---|---|
| $k_\infty$ | 1.35513 | 1.38642 | 1.41524 | 1.43321 | 1.42142 | 1.28830 | 1.21383 |
| Effect of increasing FVF on $k_\infty$ | -0.00376 | -0.00392 | -0.00279 | -0.00193 | 0.00001 | 0.00443 | 0.00494 |
| Effect of different geometries on $k_\infty$ | -0.00354 | -0.00478 | -0.00707 | -0.00769 | -0.00397 | 0.02587 | 0.04636 |
| Effect of expansion displacement on $k_\infty$ | -0.00730 | -0.00870 | -0.00986 | -0.00962 | -0.00396 | 0.03030 | 0.05130 |

**4 Conclusion:** Several graphite cell models of fuel salt channel with different sizes and shapes were established in this work, and the influence of the size and shapes of the fuel salt channel on the $k_\infty$ and the neutron energy spectrum of the cell was studied. What`s more, the variation law of the $k_\infty$ with the average chord length of the fuel salt channel under different fuel salt volume ratio and heavy metal molar ratio was calculated and analyzed. And some practical application examples of this study are illustrated in the end. The results show that, due to the variations in the size and shape of the cell, the types and times of nuclear reactions between neutrons and the material in the cell during the process from neutron generation to absorption change, resulting in the change of the physical properties of the cell. The specific conclusions are as follows:

1) Within a certain range of FVF and FC, the value of $k_\infty$ first increase and then decrease with the increase of the average chord length. And the difference of $k_\infty$ of fuel salt channel with different geometric shapes becomes larger with the increase of average chord length.

2) The neutron spectrum in the cell becomes thermal and the spectrum in the fuel salt region becomes hard with the increase of the average chord length. The neutron spectrum of flaky fuel salt channel is close to that of the circular fuel salt channel. The neutron spectrum of cylindrical fuel salt channel is similar to that of the spherical fuel salt channel. Therefore, the variation law of $k\infty$ with average chord length of the cell with flaky and circular fuel salt channel is similar, while that of the cell with cylindrical and spherical fuel salt channel is similar.

3) For the fuel salt channel cells with the same FVF and FC, $k_\infty$ of the fuel salt channels with different geometric shapes is in good agreement in the interval with average chord length less than 10. The $k_\infty$ of the fuel salt channel with different geometric shapes is greatly different in the range of the average chord length greater than 10 cm.

4) The variation law of the $k_\infty$ of cell with average chord length is universal and does not depend on FVF and FC values.

Results in this work shows the basic physical law of different geometrical fuel channel cells for molten salt reactor, which can provide reference for selection to some extent. The last part of this article has displayed some application examples of this study. In addition, relevant studies on the hydraulic and thermal properties of cells with different shape of fuel channel, such as the temperature distribution of cell and the flow field distribution of fuel channel, will be carried out in the future. Furthermore, the safety performances of the cores composed of these cells will be also verified. Ultimately, the knowledge of the neutron physics and thermal-hydraulic characteristics can provide experience and a good foundation for the core selection optimization of molten salt reactor. The results of this study have reference value for engineering application.

**Acknowledgments:** This study was supported by the Chinese TMSR Strategic Pioneer Science and Technology Project (No.XDA02010000) , the Frontier Science Key Program of Chinese Academy of Sciences (No.QYZDY-SSW-JSC016) and Shanghai Sailing Program with grant/award

Number: Y931021031.